\documentclass{PoS}

\title{J-PARC Neutrino Beamline and 1.3 MW Upgrade}

\ShortTitle{J-PARC Neutrino Beamline and 1.3 MW Upgrade}

\def\lsim{\lower.7ex\hbox{${\buildrel < \over \sim}$}}
\def\gsim{\lower.7ex\hbox{${\buildrel > \over \sim}$}}

\author{\speaker{Yuichi Oyama}\thanks{for T2K neutrino beamline group}\\
        KEK/J-PARC\\
        E-mail: \email{yuichi.oyama@kek.jp}}

\abstract{
J-PARC neutrino beamline has been operated
since 2009. Until May 2018, the maximum beam
power achieved was $\sim$500~kW.
In future, the beam power will be upgraded to 1.3~MW.
After the stable operation in late 2020s,
we can accumulate $\sim$30$\times 10^{20}$ POT data per year.
It is almost the same number as total POT accumulated in
recent 10~years.
The upgrade plan of the Main Ring as well as the
neutrino beamline is reported.
}

\FullConference{The 21st international workshop on neutrinos from accelerators (NuFact2019)\\
		August 26 - August 31, 2019\\
		Daegu, Korea}

\begin{document}

\section{Introduction}

J-PARC neutrino beamline has been operated
since 2009. Until May 2018, $31.6 \times 10^{20}$ POT (proton
of target) of 30~GeV proton beam were supplied to the T2K experiment.
The maximum beam power was $\sim$500~kW.

In future, the beam power will be upgraded to 1300~kW.
For this purpose, the Main Ring will provide 1300~kW beam to
the neutrino beamline. In addition, the neutrino
beamline must accept 1300~kW beam from the Main Ring.
The upgrade plans of the Main Ring as well as the
neutrino beamline are reported in this article.
Detailed report about the upgrade plan can be found
in the Technical Design Report (TDR)\cite{TDR}. Also some of the upgrade components were
precisely reported in other presentations \cite{yoichi,megan,kenji}.

\section{Main Ring}
The beam power is inversely proportional to the repetition cycle, and is proportional
to number of protons per pulse.
At present, the repetition cycle and number of protons per pulse are 2.48~s and
$2.6\times 10^{14}$, respectively. In future, they will be upgraded to be
1.16~s and $3.2\times 10^{14}$. 

For the change of the repetition cycle and the number of protons per pulse,
two Main Ring upgrades are in progress.
These are replacements of power supplies for main magnets and RF cavities.
Other changes such as the collimators and the fast extraction kicker
are reported in \cite{TDR}, and 
are not covered in this paper.

\subsection{Power supplies for main magnets}
For the short repetition cycle, fast energy recovery with large capacitor
banks is required. New power supplies with large capacitor banks are
under preparation. Some of the new power supplies were already
installed in the present system and have been used for the regular operation.
New power supply buildings, namely, D4, D5 and D6, were already
constructed by Japanese Fiscal Year (JFY) 2017.
All construction, installation and commissioning will be completed
by March 2022. For this purpose, a long shutdown will be needed in 2021.

\subsection{RF cavities}
Higher RF voltages are necessary both for the short repetition cycle,
and for large number of protons per pulse.
At present, 7 fundamental RF cavities and 2 second harmonic RF cavities
are used for 2.48~s repetition cycle.
For 1.16~s repetition cycle, 11 fundamental RF cavities and
2 new-type second harmonic RF cavities are needed.
New-type second harmonic RF cavities were assembled.
They will be installed during the long shutdown in 2021.
Additional fundamental RF cavities will be installed one
by one afterward.

\smallskip

After about one year shutdown in 2021, the beam power will be upgraded
to 700~kW in 2022. After 2022, additional RF magnets will be installed one by one,
and, hopefully, the beam power will be 1300~kW in 2028.  

\section{Neutrino beamline}
A detailed description about the J-PARC neutrino beamline at the beginning of
the T2K experiment can be found in \cite{NIM}. Upgrades for some of
the components are planned and are in progress. These are in the following
subsections.

\subsection{Target}

The target is made of graphite with a cylindrical shape (26~mm$\phi$ $\times$ 900~mm)
which is embedded in the first magnetic horn.
The present target was designed to accept up to 900~kW beam.

For higher beam power, the cooling system with helium gas must be changed.
The pressure of helium gas should be upgraded from 1.6~bar to $\sim$5~bar.
It improves the helium mass flow rate from 32~g/s to 60~g/s.
These estimations were obtained from simulation of the thermal stress and the cooling capacity.
The upgraded cooling system with new helium compressors is under design.
In the new design, other changes are also considered.
Specially,  safety and easiness in the remote maintenance is
a key issue in the design.

\subsection{Magnetic horns}

In the future operation, larger horn current and higher repetition cycle
are required. The horn current will be changed from 250~kA to
320~kA, and the repetition cycle must be changed from
2.48s to 1.16s in accordance with the accelerator cycle.

\begin{center}
\begin{figure}[b!]
\center{\includegraphics[width=35pc]{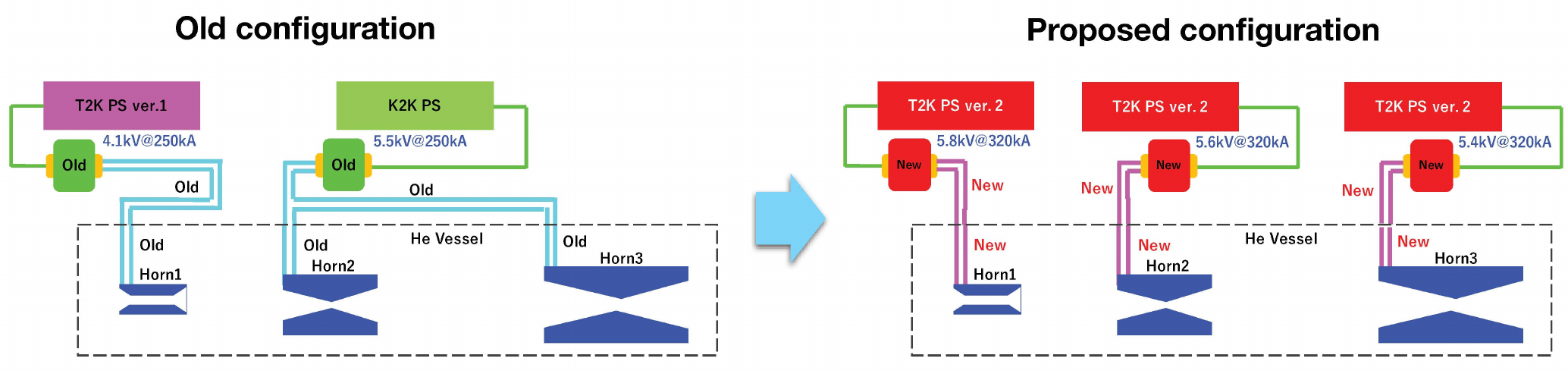}}
\caption{\label{fig:horn}
Schematic illustration of electrical system for the magnetic horns.
Present configuration (left) and future upgrade (right) are shown.
}
\end{figure}
\end{center}

For these purposes, the electrical system of the magnetic horn must
be improved. At present, electrical power for the second and the third
magnetic horn are supplied from one power supply system.
``One horn~-~one supply'' configuration (shown in Figure~\ref{fig:horn}) is
needed to satisfy above requirements.

New power supplies, new transformers, and new low impedance striplines
are already designed. To extend the cooling capacity for horn
conductors, cooling water from striplines is under consideration.

\subsection{Beam window}

The beam window separates helium vessel from vacuum in the primary beamline.
It is a double wall of 0.3~mm thick Ti-6Al-4V, cooled by helium gas (0.8~g/s).
The most serious problem is that stress by partial heat load at the beam
spot may break the window.

The current beam window can accept up to $\sim$750~kW beam.
The thickness of the beam window will be changed to > 0.4~mm to improve
the tolerance for thermal stress in the next version.

Radiation damage may change the characteristics of the thermal
stress of the material. It is an open question, and is under study as
given in the next subsection.

\subsection{Study of radiation damage}

The radiation damage of the Ti-6Al-4V has been
reported only up to 0.3$\sim$0.4~DPA (displacement per atom).
It corresponds to $~2\times 10^{20}$~POT.
When the first beam window was replaced in T2K,
$22.4\times 10^{20}$~POT had been delivered.
The radiation damage of T2K target/window were already
beyond the existing damage data.

RaDIATE International collaboration\cite{RaDIATE} is studying
the effect of radiation damage on target/window
materials since 2012.
J-PARC officially joined the project in 2017.
Irradiation of various material samples at BLIP (BNL) have been studied.
Use of other materials for beam window and other beamline components is
under investigation based on the experimental data.

\subsection{Cooling water for helium vessel, decay volume and beam dump}

Large fractions of the beam energy are absorbed in the helium vessel, the decay volume and
the beam dump. 
Thermal stress may damage the structure of the
helium vessel and the decay volume.
Their temperature must be kept below 60~$^{\circ}$C.
The temperature of the beam dump core must be below 400~$^{\circ}$C. 
They are cooled by cooling water system.
At present, the system is adjusted for the 1000~kW operation.
The cooling power can be upgraded by increasing flow rate of the cooling water.
Some components in the cooling water circulation system must be exchanged or added. They are circulation pumps, heat exchangers, chillers and cooling towers.
Radioactivity of the cooling water will become higher.
This is another serious problem. 

\subsection{Drainage of radio-active water}

About 20~m$^{3}$ of cooling water in the neutrino beam line is exposed
to neutrons from the proton beam, and are highly activated.
Active water are stored in a $\sim$20~m$^{3}$ of a buffer tank.
Then they are sent to two 50~m$^{3}$ disposal tanks and diluted to satisfy tritium
concentration < 42~Bq/cc,
which is required from a regulation for drainage.
The drainage can be done every three business day cycle.
Present disposal capability is $\sim$180~GBq/year of tritium.
However, $\sim$600~GBq/year of tritium are produced for 1300~kW beam.
The present disposal tanks are too small, and we are planning to construct
two $\sim$200~m$^{3}$ disposal tanks. 

\section{Summary and outlook}

\begin{center}
\begin{figure}[t!]
\center{\includegraphics[width=25pc]{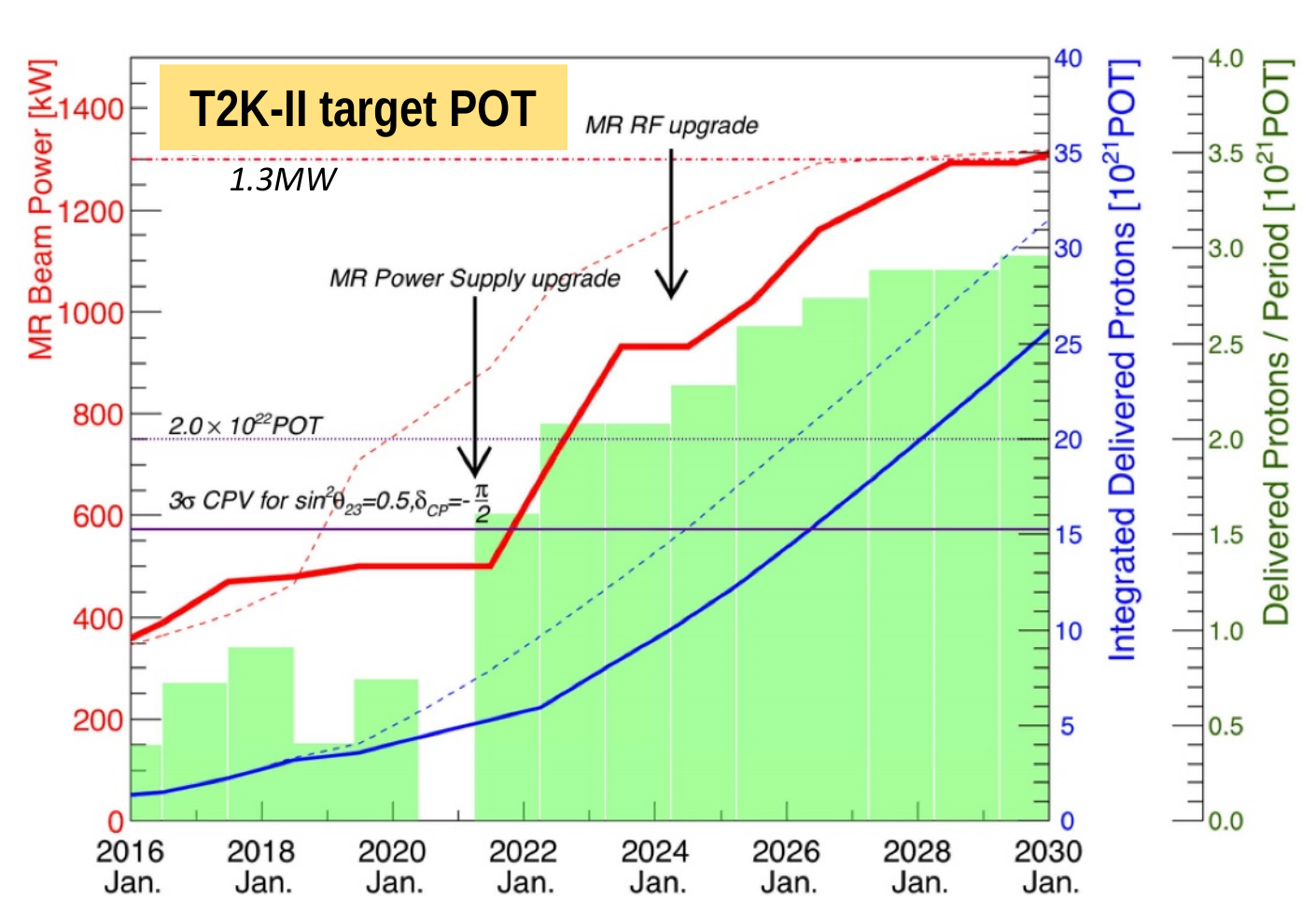}}
\caption{\label{fig:targetpot}
The target MR beam power and accumulated POT as a function of Japanese Fiscal Year
(JFY). The solid lines are the target MR beam power (red) and accumulated POT (blue) where six months of the MR operation with the fast-extraction mode each year and the running time efficiency of 90\% are assumed.
}
\end{figure}
\end{center}

All upgrades in the neutrino beamline will be finished by Mar. 2022.
The beamline will wait for delivery of 1300~kW beam from the Main Ring.
The target, magnetic horns and beam windows need periodical exchange.  Additional upgrades will be
needed at the exchanges after 2022.
After 1300~kW beam operation become stable,
we can accumulate $\sim 30\times 10^{20}$ POT data per year
as shown in Figure~\ref{fig:targetpot}. It is almost the same number as total
POT accumulated in $\sim$10~years until 2019.

\end{document}